\documentclass[aps,pra,twocolumn,superscriptaddress]{revtex4-2}
\pdfoutput=1
\usepackage{graphicx}
\usepackage{amssymb,amsmath,amsthm}
\usepackage{bm,bbm}
\usepackage{array}
\usepackage[bookmarks=false]{hyperref}
\hypersetup{colorlinks=true,citecolor=blue,linkcolor=red,%
	urlcolor=blue,pdfstartview=FitH,bookmarksopen=true}
\makeatother

\newcommand{\Rmnum}[1]{\expandafter\@slowromancap\romannumeral #1@}
\newcommand{\kb}[2]{\vert #1 \rangle \langle #2 \vert}
\newcommand{\kt}[1]{\vert #1 \rangle}

\usepackage{xcolor}

\begin{document}

	\title{One-way Einstein--Podolsky--Rosen steering beyond qubits}
	
	\author{Qiang Zeng}
	\email{zengqiang@baqis.ac.cn}
	\affiliation{Beijing Academy of Quantum Information Sciences, Beijing 100193, China}
	\affiliation{Key Laboratory of Advanced Optoelectronic Quantum Architecture and Measurement of Ministry of Education, School of Physics, Beijing Institute of Technology, Beijing 100081, China}
	\affiliation{Centre for Quantum Computation and Communication Technology (Australian Research Council), Centre for Quantum Dynamics, Griffith University, Brisbane, Queensland 4111, Australia}
	
	\date{\today}

\begin{abstract}
	Quantum steering has been exploited as an important resource in modern quantum information processing.
	Owing to its directional nature, some quantum states that are asymmetric under the exchange of parties have been found to manifest steering only in specific direction, thus called one-way steering.
	Existing works focused on one-way steering in systems of qubits. 
	Here we propose a family of two-party states that are one-way steerable in systems of $d$-dimension.
	In particular, we validate the one-way steerability of the states for $d=3$, and demonstrate how one-way steering parameter space manifests in two-qutrit system. 
	A general numerical approach for characterizing higher-dimensional one-way steering is provided.
	Moreover, we develop a method to characterize one-way steering with the experimental loss taken into account, with which the tradeoff relation between losses and measurement settings in steering test in higher-dimensional system is investigated.
	Our loss-counted model works for finite-dimensional system with finite measurement settings.					
\end{abstract}

	\maketitle
	
\section{Introduction}
	Quantum entanglement, which has a clear mathematical structure (that is, the state of the system is non-factorizable), is seen as a counter-intuitive effect exemplified by real physical systems.
	This led to the dispute on the completeness of quantum mechanics that was raised from the famous Einstein--Podolsky--Rosen argument~\cite{epr}, and the dispute was later reformed into the discussion and certification of the violation of Bell inequalities~\cite{bell1964Einstein}.
	Indeed, it is not revealed until recently that quantum entanglement, in the view of correlation, has far richer structure between non-factorizable and Bell-nonlocal~\cite{wiseman2007Steering,saunders2010Experimental,quintino2015Inequivalence}.
	
	The intermediate form of entanglement structure is named \textit{steering}~\cite{uola2020Quantum}, which describes the effect that one of the entangled parties, by performing local measurements, can affect the local state of another remote party.
	Different from the other two entanglement structures, quantum steering is defined in the asymmetric form, as the \textit{steering} and the \textit{steered parties} are distinct and cannot be interchanged.
	To capture the essence of quantum steering, one can interpret it in the view of theoretical information tasks, where steering is demonstrated if the steered ensemble cannot be explained by a local hidden state (LHS) model~\cite{jones2007Entanglement}.
	
	As steering has direct correspondence to information tasks, various applications of steering have been found, such as one-sided device-independent quantum key distribution~\cite{branciard2012Onesided}, randomness certification~\cite{skrzypczyk2018Maximal,guo2019Experimental}, subchannel discrimination~\cite{piani2009All}, device-independent quantification of measurement incompatibility~\cite{cavalcanti2016Quantitative}, secure quantum teleportation~\cite{reid2013Signifying} and secret sharing~\cite{xiang2017Multipartite}. 
	
	In particular, the asymmetry of steering plays a pivotal role in those applications, which manifests in the different levels of trust to the involved parties~\cite{cavalcanti2013Entanglement}. 
	It has been shown that the detection loophole at the untrusted side can be effectively closed using loss-tolerant steering criteria, thus promising information-theoretical security~\cite{bennet2012Arbitrarily}.
	However, the asymmetry of steering does not only manifest when imposed different levels of trust to parties, which should be addressed as the directionality of steering~\cite{FurtherExplain}, but rather with the state \textit{per~se} is of an asymmetric form, in which case the steering is unidirectional---known as one-way steering.
	
	One-way steering, as identified in 2007~\cite{wiseman2007Steering}, was first studied in the Gaussian regime~\cite{reid1989Demonstration,midgley2010Asymmetric,Wagner541,handchen2012Observation}, and was subsequently studied in discrete variable regime.
	Specifically, two-qubit states in tailored forms were found to exhibit one-way steering, by using projective measurements~\cite{bowles2014Oneway} and positive-operator-valued measures (POVMs)~\cite{quintino2015Inequivalence} respectively. 
	Further, a family of two-parameter mixed states that exhibit one-way steering are characterized~\cite{bowles2016Sufficient}.
	Recently, sufficient conditions for certifying one-way steering of two-qubit system have also been further investigated~\cite{baker2018Necessary,baker2020Necessary}.
	
	Following the theoretical works, some experimental demonstrations of one-way steering have been carried out. 
	It was first observed using two-setting projective measurements~\cite{sun2016Experimental}, and later extended to using multisetting projective measurements~\cite{xiao2017Demonstration}.
	One-way steering with POVMs has also been demonstrated~\cite{wollmann2016Observation}.
	Recently, conclusive one-way steering with infinite number of measurements was first demonstrated for qubit-qutrit system (or two-qubit with one of them has a third outcome of vacuum state)~\cite{tischler2018Conclusive}, and later was demonstrated reducing to the genuine two-qubit system~\cite{zeng2020reliable,nguyen2019Geometry}.
	
	However, existing works on one-way steering focused on two-qubit or qubit-qutrit systems, and the genuine high-dimensional one-way steering has not been studied to date. 
	Indeed, steering that is beyond qubit system is of special interests, since high-dimensional steering possesses strong robustness against noise, which has been theoretically predicted and experimentally demonstrated~\cite{zeng2018}.
		
	In this work, we extend the exploration of one-way steering to genuine high-dimensional systems. 
	Specifically, we first introduce the $d$-dimensional partially entangled states, which is the generalization of its two-qubit form.
	Then, taking three-dimensional partially entangled states for example, we illustrate how one-way steering manifests in high-dimensional systems.
	Next, we propose a loss-counted model for characterizing $d$-dimensional EPR steering in a more practical manner, and accordingly develop an operational numerical method. 
	Finally, we apply our method to characterizing quantitatively the relation between the measurement settings and detection efficiencies of two-qutrit one-way steering states.
	
	This article is organized as follows: 
	In Sec.~\ref{II} we briefly review the LHS model and propose the $d$-dimensional partially entangled states. 
	In Sec.~\ref{III} we characterize the one-way steerability of the proposed states for $d=3$ based on numerical approach with respect to mutually unbiased bases measurements and further the general measurements. 
	In Sec.~\ref{IV} we introduce our loss-counted model and apply it to high-dimensional two-way steering states (in comparison with known analytical results) and two-qutrit one-way steering states respectively.
	We summarize the main results and discuss potential applications of the work in Sec.~\ref{V}.
	
\section{Partially entangled states}\label{II}
	Consider the two-party system, where Alice and Bob are sharing a quantum state $\rho_{AB}$. 
	If Alice performs local measurements on her subsystem, which are denoted by a set of operators $\{M_{a|x}\}_{a,x}$, with $M_{a|x} \ge 0$ and $\sum_a M_{a|x}=\mathbbm{1}$, where $x$ denotes her measurement setting and $a$ the corresponding outcome, then on Bob's subsystem, conditioned on the measurement $x$ and outcome $a$, a collection of subnormalized density matrices $\{\sigma_{a|x}\}_{a,x}$ are generated, with 
	\begin{equation}\label{}
		\sigma_{a|x}=\text{Tr}_A(M_{a|x}\otimes \mathbbm{1} \rho_{AB}).
	\end{equation}
	This set of matrices is called \textit{assemblage}~\cite{pusey2013Negativity}, and is proven to be useful in characterizing the steerability from Alice to Bob (or vice versa if measurements are performed on Bob's side).
	Specifically, the assemblage is called unsteerable if it can be reproduced by a LHS model~\cite{wiseman2007Steering}, namely, it can be decomposed into 
	\begin{equation}\label{LHS}
	\sigma^{\text{LHS}}_{a|x}=\int \sigma_{\lambda}q(\lambda) p(a|x,\lambda)d\lambda,~\forall a,x,
	\end{equation}
	where $\{\sigma_\lambda\}$ is a set of positive matrices with some probability distribution $q(\lambda)p(a|x,\lambda)$.
	Any assemblage that refutes a LHS model demonstrates EPR-steering.
	Interestingly, assemblages generated from entangled states do not guarantee steering.
	Indeed, given an assemblage, certifying whether it admits a LHS model is difficult, not mention to one has to consider the experimental imperfections in practical demonstration.
\begin{figure}[t]
	\includegraphics[width=\columnwidth]{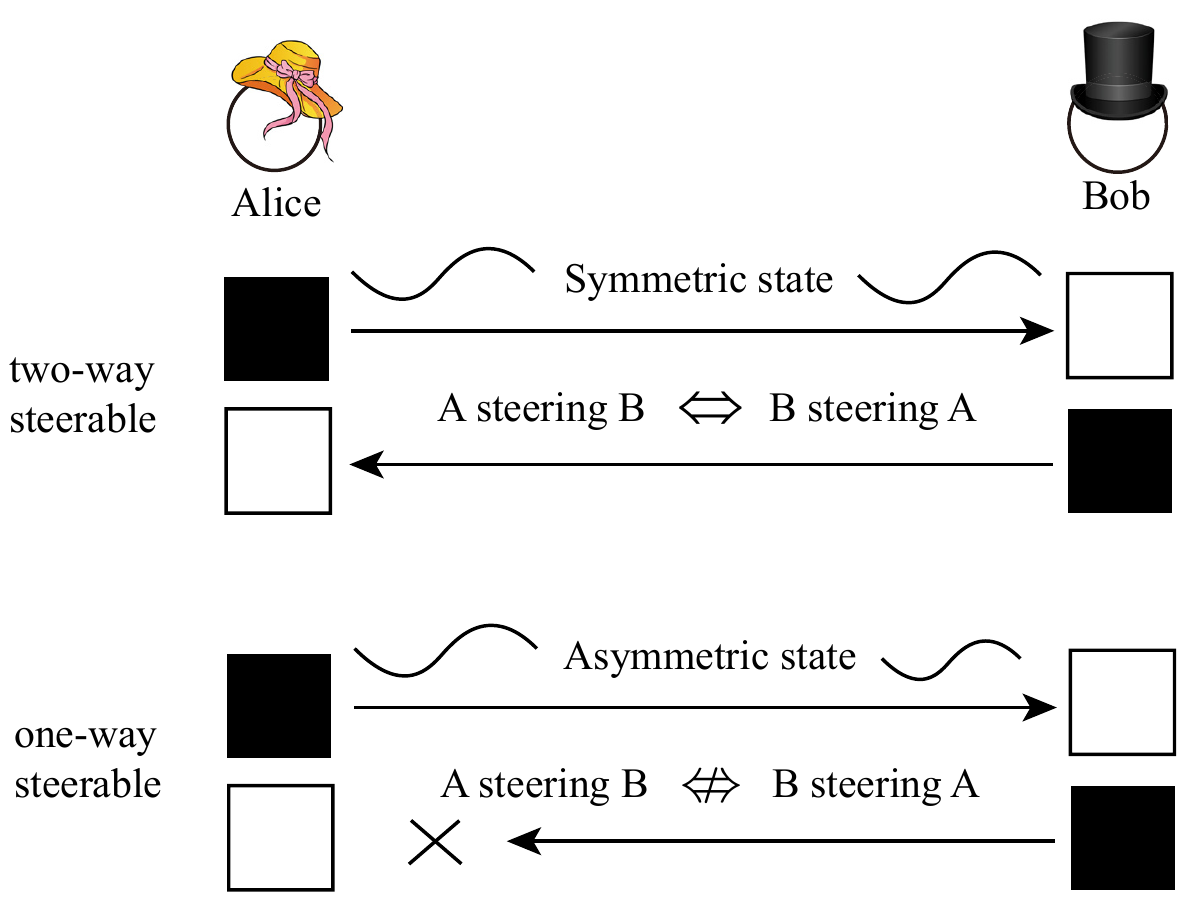}
	\caption{\label{fig1} 
		The correspondence of symmetric states to two-way steering and asymmetric states to one-way steering. 
		Note here the symmetric state does not only refer to the state with symmetric form, but rather the ones can demonstrate steering under the exchange of the steering and steered parties.
		The black squares represent the uncharacterized devices, while the white squares denote the trusted devices.
	}
\end{figure}

	It can be seen that the steering party and the steered party have already been designated for an assemblage, given a quantum state and the local measurements. 
	Thus certifying steering is in itself directional, even if the given quantum state has a symmetric form.
	Traditionally, people use symmetric states to study steering, and the conclusion can be straightforwardly deduced when the role of two parties are interchanged.
	
	As illustrated in Fig.~\ref{fig1}, for the symmetric states, if Alice is able to steer Bob, then Bob must be able to steer Alice.
	While if the states are asymmetric under the exchange of parties, then the conclusions may differ for different roles of the parties.
	It is pointed out that one-way steering can only occur for mixed states, as the pure entangled states can always be transformed to a symmetric form by local basis change~\cite{bowles2014Oneway}.
	
	Recently, a family of mixed states with two independent parameters have been found, which are shown to be one-way steerable for specific combination of parameters.
	This family of states are called two-qubit partially entangled state~\cite{bowles2016Sufficient}, with
	\begin{equation}
		\rho(p,\theta)=p\kb{\psi_\theta}{\psi_\theta}+(1-p)\rho_{\theta}^{A}\otimes \mathbbm{1}/2,
	\end{equation}
	where $\kt{\psi_\theta}=\text{cos}(\theta)\kt{00}+\text{sin}(\theta)\kt{11}$, $ \rho_{\theta}^{A}=\text{Tr}_B\kb{\psi_\theta}{\psi_\theta}$, $p \in [0,1]$ and $\theta \in [0,\pi/4]$.
	
	Here we generalize the above states to its high-dimensional form, i.e.,
	\begin{equation}\label{dpes}
		\rho(p,\bm{a})=p\kb{\psi^{+}_{\bm{a}}}{\psi^{+}_{\bm{a}}}
		+(1-p)\rho_{\bm{a}}^{A}\otimes \mathbbm{1}/d,
	\end{equation}
	where $\kt{\psi^+_{\bm{a}}}=\sum_{i=1}^{d} a_i \kt{ii}$, $\rho_{\bm{a}}^{A}=\text{Tr}_B\kb{\psi^{+}_{\bm{a}}}{\psi^{+}_{\bm{a}}}$, $p \in [0,1]$, $d$ is the dimension of the system and $\bm{a}$ is a $d$-dimensional unit vector.

\section{One-way steering of two-qutrit state}\label{III}
\subsection{Mutually unbiased bases measurements}
	With the asymmetric states being targeted, we start to consider the way to characterize the steerability of the state in both directions, respectively.
	Indeed, one finds that characterizing steerability for infinite measurement settings is tricky, since the variable $\lambda$ in Eq.~\eqref{LHS} could take infinitely many values~\cite{cavalcanti2017Quantum}.
	However, if we limit the number of measurements and outputs to finite values, the problem would become computationally feasible~\cite{skrzypczyk2014Quantifying,piani2015Necessary}.
	In particular, if we further adopt the ${d+1}$ mutually unbiased bases (MUB) as the measurements on Alice's side---let's consider Alice to steer Bob first, Alice performs ${d+1}$ measurements and obtains $d$ outcomes.
	
	For each of the conditional state $\sigma_{a|x}$ on Bob's side, where $a=\{0,...,{d-1}\}$ and $x=\{0,...,d\}$, we are interested in whether the following equation holds
	\begin{equation}\label{siglam}
		\sigma_{a|x}=\sum_{\lambda=0}^\Lambda D(a|x,\lambda)\sigma_{\lambda},
	\end{equation}
	where $\Lambda=d^{({d+1})}-1$, indicating the number of local states that required to construct the LHS model. Here $D(a|x,\lambda)$ is the deterministic single-party conditional probability distribution, which gives a fixed outcome $a$ for each measurement $x$.
	
	It is developed and has been widely acknowledged that the above problem can be fully characterized using semidefinite programming (SDP). Indeed, the SDP approach defines the measure of steering called \textit{steering weight} (SW)~\cite{skrzypczyk2014Quantifying}, which can be seen as the linear proportion of the steerable part of the given assemblage, i.e.,
	\begin{equation}\label{propotion}
		\sigma_{a|x}=\mu\sigma_{a|x}^{\text{LHS}}+(1-\mu)\sigma_{a|x}^{\text{ST}},~\forall a,x,~0 \le\mu\le1,	
	\end{equation}
	where $1-\text{max}(\mu)$ is defined as the steering weight, and the maximization of $\mu$ can be computed via the SDP defined in Ref.~\cite{skrzypczyk2014Quantifying}.
	
\begin{figure}[t]
	\includegraphics[width=\columnwidth]{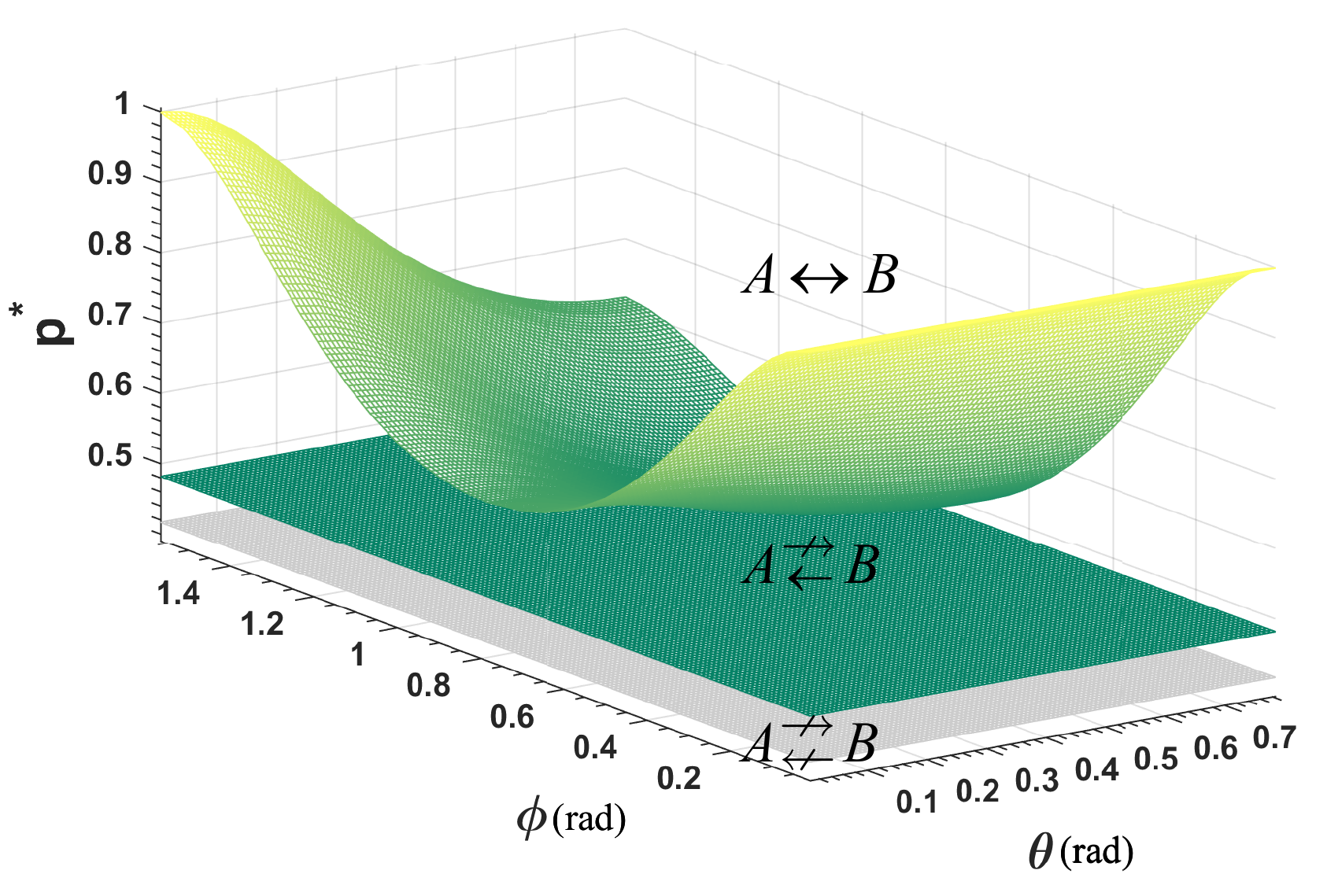}
	\caption{\label{fig2}  One-way steering demonstration of two-qutrit partially entangled state using four-setting mutually unbiased bases.
		The upper surface denotes the critical value of parameter $p$ given various $\theta$ and $\phi$ considering Alice to steer Bob.
		The lower plane corresponds to the results for Bob to steer Alice.
		The strict lower bound of $p^*_{B\rightarrow A}$ given general measurements is also presented at the lowest in gray for reference.
	}
\end{figure}

	Take the two-qutrit partially entangled states for example, in the following we show in detail how one-way steering manifests in the high-dimensional system.
	Specifically, the two-qutrit partially entangled states forms as
	\begin{equation}
		\rho(p,\theta,\phi)=p\kb{\psi^{+}_{\theta,\phi}}{\psi^{+}_{\theta,\phi}}+(1-p)\rho_{\theta,\phi}^{A}\otimes \mathbbm{1}/3,
	\end{equation}
	where $\kt{\psi^{+}_{\theta,\phi}}=\text{cos}(\theta)\text{sin}(\phi)\kt{00}
	+\text{sin}(\theta)\text{sin}(\phi)\kt{11}
	+\text{cos}(\phi)\kt{22}$, $p \in [0,1]$ $\theta \in [0,\pi/4]$ and $\phi \in [0,\pi/2]$.
	In Fig.~\ref{fig2}, we present the distribution of the three parameters of two-qutrit partially entangled states demonstrating one-way steering.
	Here we use $p^*$ to work as the measure of steering, which gives the critical value upon which steering weight vanishes.
	The upper surface shows the critical value $p^*$ given certain $\theta$ and $\phi$, when considering Alice to steer Bob.
	Below the surface, the steering weight is non-positive, which indicates non-steerability.
	Only if the parameter $p$ is greater than $p^*$, which corresponds to the space above the surface, steering from Alice to Bob is certified.

	On the other hand, the critical value for Bob to steer Alice turns out to form a plane, which means Bob can always steer Alice, as long as the proportion of the maximally entangled state is greater than 0.4818 (within numerical precision), and is irrelevant to the other two parameters.
	Therefore we conclude that the partially entangled states with parameters that are in the space above the plane and below the surface demonstrate one-way steering.

\subsection{General measurements}	
	The above results are based on four-setting MUB (see results and discussions for two- and three-setting cases in Appendix~\ref{S:I}).
	Indeed, it is known that to conclusively characterize one-way steering, one needs to consider general measurements, since to certify non-steerability, it is required to verify if all the assemblages generated from the state admit a LHS model. 
	If, for example, increasing the number of measurement settings, it is shown both theoretically~\cite{saunders2010Experimental,evans2013Losstolerant,evans2014Optimal} and experimentally~\cite{bennet2012Arbitrarily} that one can turn non-steerability into steerability.
	The intuitive reason for this is that the extra measurements brings more terms of constraints to the LHS model in Eq.~\eqref{siglam}, thus limits the LHS model of $\{\sigma_{a|x}^{\text{LHS}}\}$ in Eq.~\eqref{propotion} in reproducing the assemblage $\{\sigma_{a|x}\}$, which leads to the increased steering weight value. 
	
	For $d$-dimensional partially entangled state, it has been  proven analytically~\cite{jones2007Entanglement} that the strict lower bound of the critical value considering Bob to steer Alice is $p^*_{B\rightarrow A}(d)=(H_d-1)/(d-1)$, where $H_d=1+1/2+\cdots+1/d$.
	This follows the idea that one can always transform the partially entangled state to the $d$-dimensional isotropic state via some filtering operation on Alice~\cite{quintino2015Inequivalence}, without affecting the steerability from Bob to Alice~\cite{bowles2016Sufficient}.
	
	Certifying the non-steerability in the opposite direction is rather difficult.
	Indeed, the authors in Refs.~\cite{cavalcanti2016General,hirsch2016Algorithmic} proposed the general methods for constructing the LHS model for given states in an asymptotically way.
	Considering $p^*_{B\rightarrow A}$ is independent of $\theta$ and $\phi$, using the aforementioned methods, one may expect that $d$-dimensional partially entangled states can always demonstrate one-way steering for certain combination of parameters, even when it comes to general measurements.
	
	The main insight in finding the lower bound of $p^*_{A\rightarrow B}$ with respect to general measurements is to note that applying noisy measurements on a state (which could even be assumed non-physical), is equivalent, at the level of state preparation for Bob, to applying noise-free measurements on a noisy version of the state.
	Thus if the state admits a LHS model for the noisy measurements, then the noisy version of the state must admit a LHS model for the noise-free measurements.
	In qubit case specifically, one could start with the noisy measurements $M^{\eta}_{\pm|\hat v}=\left[\mathbbm{1}\pm\eta(\hat v\cdot \vec{\sigma})\right]/2$ with $\eta=0.79$, where $\vec{\sigma}$ are Pauli matrices, $\hat v$ gives the direction of Bloch vectors and 0.79 corresponds to the radius of the inscribed sphere of the icosahedron fitting inside the Bloch sphere.
	It is known that Werner state with visibility $V\lesssim0.54$ admits a LHS model for the noisy measurements $M^{\eta=0.79}_{\pm|\hat v}$~\cite{hirsch2016Algorithmic}, and thus its noisy version with $V\lesssim0.54*0.79=0.43$ admits a LHS model for all projective measurements.
	
	Following the terminology in Refs.~\cite{cavalcanti2016General,hirsch2016Algorithmic}, we call $\eta$ the shrinking factor, which indicates to what extent the given noisy measurements could represent the general measurements in characterizing non-steerability.
	Then we solve the following SDP problem, which is the relaxation of the feasibility SDP presented in Refs.~\cite{cavalcanti2016General,hirsch2016Algorithmic}, to detect the lower bound of $p^*_{A\rightarrow B}$ given general measurements:
	\begin{align}
		\max~~&Q=\text{Tr}\sum_\lambda \sigma_\lambda\\ \nonumber
		\text{over}~~&O_{AB},~\sigma_\lambda\\ \nonumber
		\text{s.~t.}~~&\text{Tr}_A(M_{a|x}\otimes \mathbbm{1} O_{AB})=\sum_\lambda D(a|x,\lambda)\sigma_{\lambda},~\forall a,x,\lambda\\
		&\eta O_{AB}+(1-\eta)\rho^A\otimes O_B=\rho_{AB},	\nonumber	
	\end{align}
	where $O_{AB}$ is any hermitian operator, $\{M_{a|x}\}$ is the $d+1$ MUB, $\{D(a|x)\}$ is the deterministic distribution of the $d+1$ MUB and $\eta$ corresponds to the ratio between the radius of the inscribed sphere of the polytope formed by the $d+1$ MUB to the radius of the generalized unit sphere in $d^2-1$ dimensional space, which according to Ref.~\cite{bengtsson2005Mutually} is $1/\sqrt{(d^2-1)(d-1)}$.
	Note that this result applies only for prime-power dimension, and finding the optimal shrinking factor is always a tricky problem, even for 2-dimensional system---increasing the shrinking factor does not necessarily give a tighter bound of non-steerability~\cite{hirsch2016Algorithmic}.
	
	Similar to finding the critical value $p^*$ in the previous case, instead of using the vanishing of steering weight, here we find $p^*$ upon $Q=1$, and $\rho_{AB}$ is sufficiently unsteerable to general measurements if $p\le p^*$.
\begin{figure}[t]
	\includegraphics[width=\columnwidth]{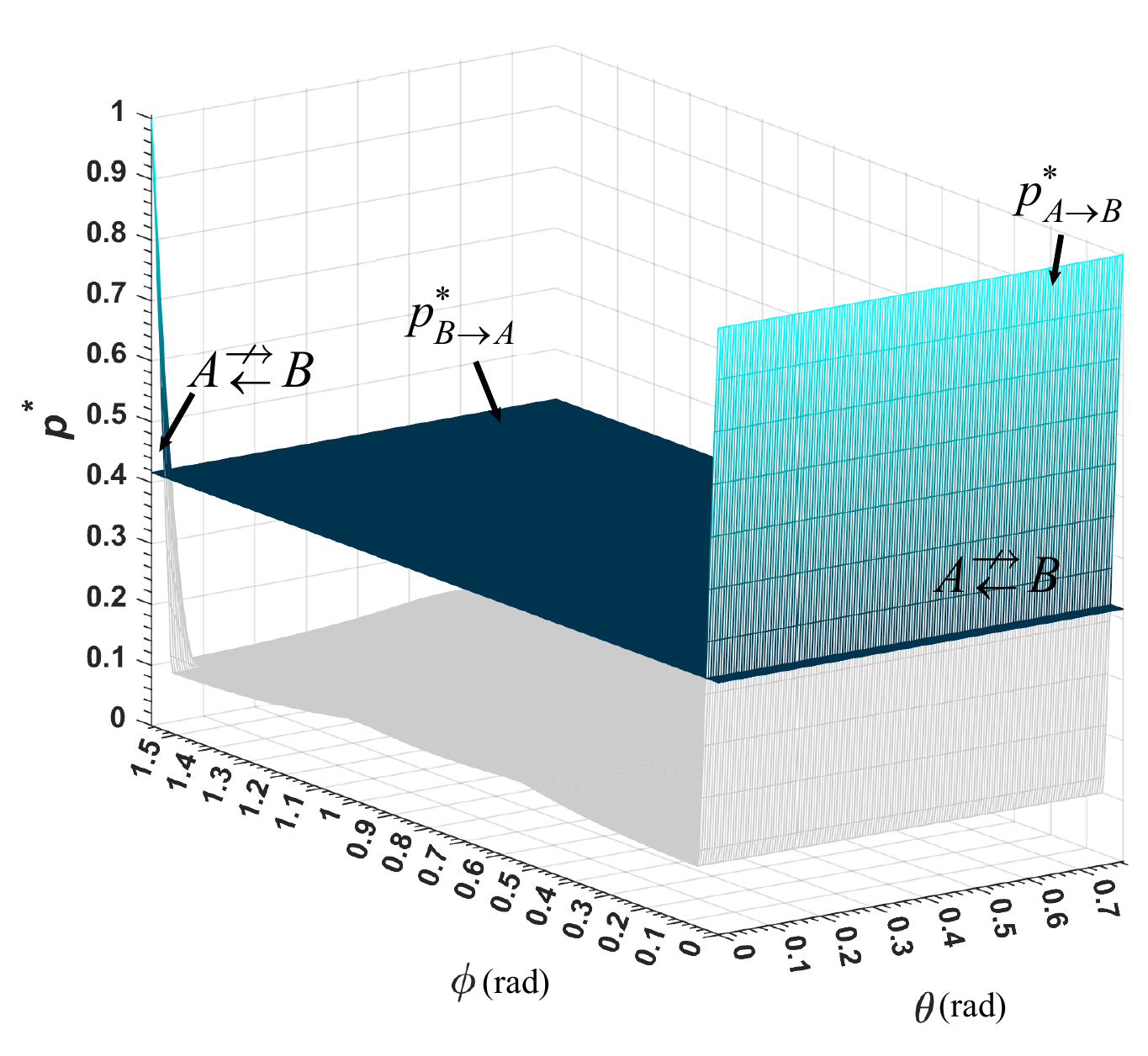}
	\caption{\label{Fig3}  
		Demonstration of one-way steering of two-qutrit state with respect to general measurements.
		For those states with parameters are in the space above the plane and below the curved surface, the one-way steering effect is conclusively certified, in the sense that the general measurements are considered in both directions.
		The curved surface with $p^*<p^*_{B\rightarrow A}$ is painted in gray, for the steerability of the states with parameters in such space is uncertain using current SDP settings.
		Note that this uncertain space is not the same as the steerability-reversing space appearing in the two-setting MUB case in Appendix~\ref{S:I}.
	}
\end{figure}
	Again take two-qutrit case for example, we compute the distribution of $p^*_{A\rightarrow B}$, and the results are shown in Fig.~\ref{Fig3}.
	The curved surface ranging from approximately 0.1 to 1 corresponds to the computed $p^*_{A\rightarrow B}$, while the plane corresponds to the strict lower bound $p^*_{B\rightarrow A}$.
	
	It is worth noting that the computed $p^*_{A\rightarrow B}$ only gives a sufficient bound for the non-steerability, and could be significantly improved if considering more refined measurement settings and the corresponding shrinking factors.
	Despite this, one still finds that two-qutrit partially entangled states are of distinct one-way steering zone, which indicates that it can always demonstrate one-way steering for certain combination of parameters, even when the steering party implementing general measurements. 
	
	For the higher-dimensional cases (of prime-power number, if adopting the MUB shrinking factor formula), we note that the SDP could become dramatically resource-demanding. 
	Nevertheless, the method provides an elementary solution to estimating a reliable lower bound of non-steerability.

\section{Loss-counted one-way steering}\label{IV}
\subsection{Loss-counted model}
	To conclude the non-steerability for a given state, one in principle is required to test infinite-setting measurements, which (without provoking ancillary dimension~\cite{wollmann2016Observation,tischler2018Conclusive}) is not feasible in the experiment.
	On the other hand, there is another prominent factor that affects one-way steering demonstration, which is the inevitable experimental losses.
	In photonic experiment for example, loss happens naturally, which manifests in no-click event being registered at the measurement devices.
	This should not be a problem if the device is trusted.
	
	However, if the steering party, say Alice, is holding a cracked device, her measurement outcomes could be deliberately discarded, whenever the measurements she performs do not correspond to Bob's announced measurements, thus faking the perfect correlations and passing the steering test.
	This opens the so-called \textit{detection loophole}~\cite{pearle1970HiddenVariable} which is also significant in Bell-nonlocality test~\cite{branciard2011Detection}.
	
	From the view of LHS model, the discarded results grant more flexibility to the combination of local model in Eq.~\eqref{siglam}, which leads to the decreased steering weight value, thus indicating that steering is demolished---or equivalently, the criterion of demonstrating steering is pushed up.
	Indeed, the tradeoff relation between the losses (which is commonly characterized by the \textit{heralding efficiency} $\epsilon$, describing the probability that the steering party heralds the result by declaring a non-null prediction) and the number of measurement settings has been investigated in two-qubit system~\cite{evans2013Losstolerant,bennet2012Arbitrarily}.
	 
	Here, we aim to develop a quantitative method to characterize the relation between the heralding efficiency and the number of measurement settings in the steering test for any finite-dimensional system.
	Then, we apply our method to the one-way steering characterization.
	The influence of losses in the steering test can be interpreted as the extra outcome for each measurement setting~\cite{skrzypczyk2015Losstolerant,sainz2016Adjusting}.
	Take the $d+1$ MUB measurement for example, Alice performs $d+1$ measurements, and in the loss-counted scenario, she obtains $d+1$ outcomes for each measurement setting.
	
	We denote the extra outcome which corresponds to no-event being registered by $a=d$, as for now $a=\{0,...,d\}$.
	The assemblage in loss-counted scenario is called \textit{priori assemblage}~\cite{branciard2011Detection}, which we denote as $\{\sigma^{\text{pri}}_{a|x}\}$.
	
	With the priori assemblage, we naturally have 
	\begin{equation*} 
		\sigma^{\text{pri}}_{a|x}= \left \{ 
			\begin{array}{ll}
			 	\epsilon \sigma_{a|x}, & \forall a,x~\text{with}~a \ne d, \\ 
			 	(1-\epsilon) \rho^B, & \forall a,x~\text{with}~a=d,
		 	\end{array} 
 		\right.
	\end{equation*}
	for the probability conservation, where $\rho^B$ is the reduced state of Bob.
	With this decomposition of $\{\sigma^{\text{pri}}_{a|x}\}$, one straightforward way to understand the loss-counted scenario is to consider Alice as the distributor, and she with some probability sends Bob a mixed state that will not be discerned by Bob.
	In the meantime, she claims that those outcomes from mixed state are lost (discarded), then she can convince Bob that she is indeed able to steer his state---we assume that $\{\sigma_{a|x}\}$ is a steerable assemblage---at a relatively low cost (with a discount of $\epsilon$), since preparing a steerable assemblage $\{\sigma_{a|x}\}$ is obviously harder than preparing a local mixed state $\rho^B$.
	
	With this in mind, a decent way for Bob to deal with the losses is to assume that all the losses result from the trick of mixed state, but still believe that the rest of the experimental results are conclusive.
	Thus the LHS model in loss-counted scenario can be reformed as 
	\begin{equation}\label{prisiglam}
	  \begin{array}{rl}
		\sigma^{\text{LHS}}_{a|x}=\frac{1}{\epsilon}\sum_{\lambda=0}^\Lambda D(a|x,\lambda)\sigma_{\lambda}, & \forall a,x~\text{with}~a \ne d, \\ 
		\rho^B=\frac{1}{1-\epsilon}\sum_{\lambda=0}^\Lambda D(a|x,\lambda)\sigma_{\lambda}, & \forall a,x~\text{with}~a = d, 
	  \end{array}
	\end{equation}
	where $\Lambda=({d+1})^{({d+1})}-1$.
\begin{figure}[t]
	\includegraphics[width=\columnwidth]{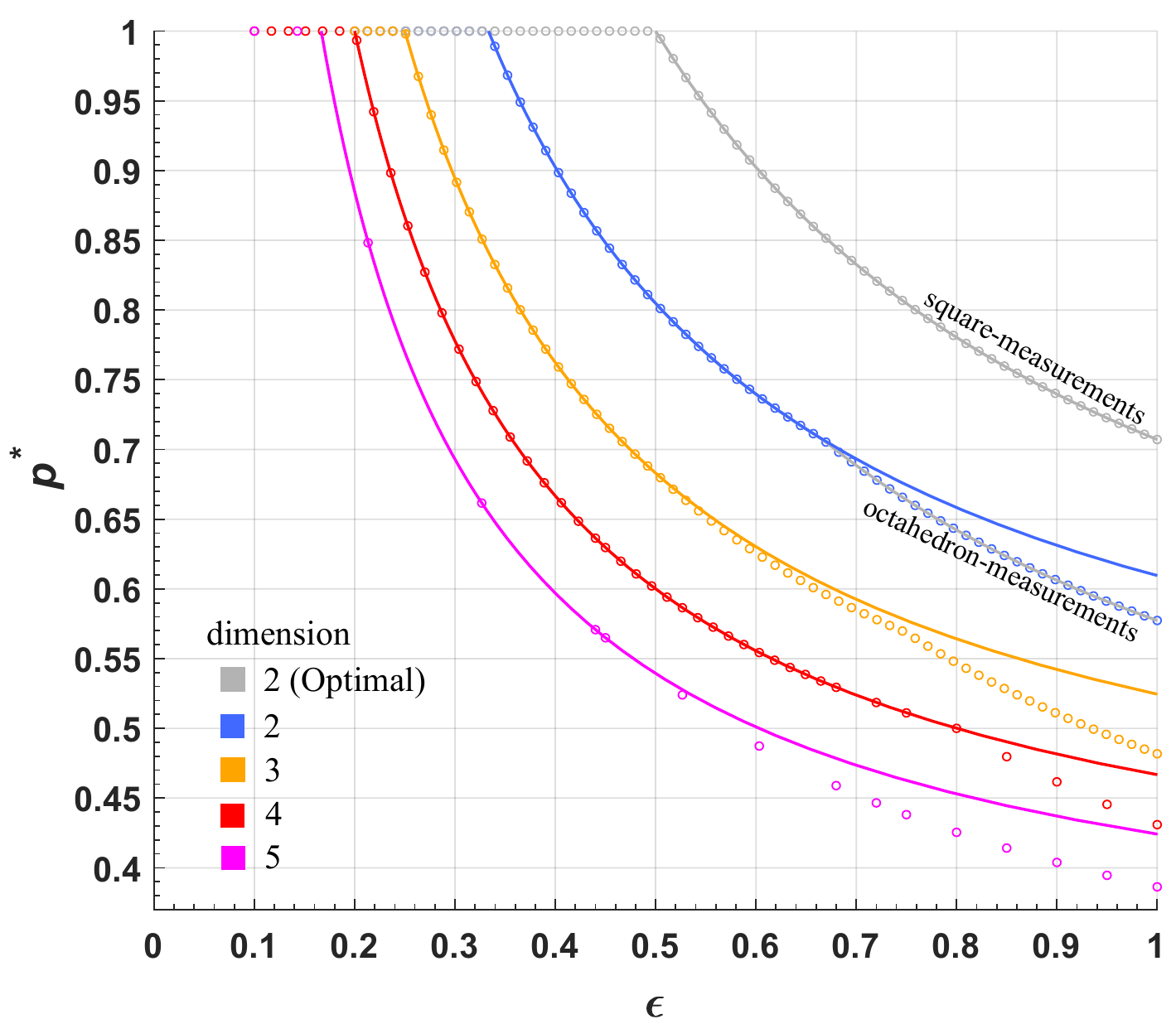}
	\caption{\label{Fig4}  
		Comparison of loss-counted model and known analytical bounds.
		The circles denote our numerical results for specific dimension, measurement setting and heralding efficiency.
		The colored curves correspond to bounds derived in Ref.~\cite{skrzypczyk2015Losstolerant} with ${d+1}$ MUB.
		Specifically, the six curves from top to bottom correspond to dimension two to five respectively.
		Note that the two gray curves at the top (specified in the figure) correspond to the optimal bounds derived in Ref.~\cite{evans2014Optimal}, which are for two-qubit system with square- and octahedron-Platonic solid measurements respectively.
	}
\end{figure}
\begin{figure}[t]
	\includegraphics[width=\columnwidth]{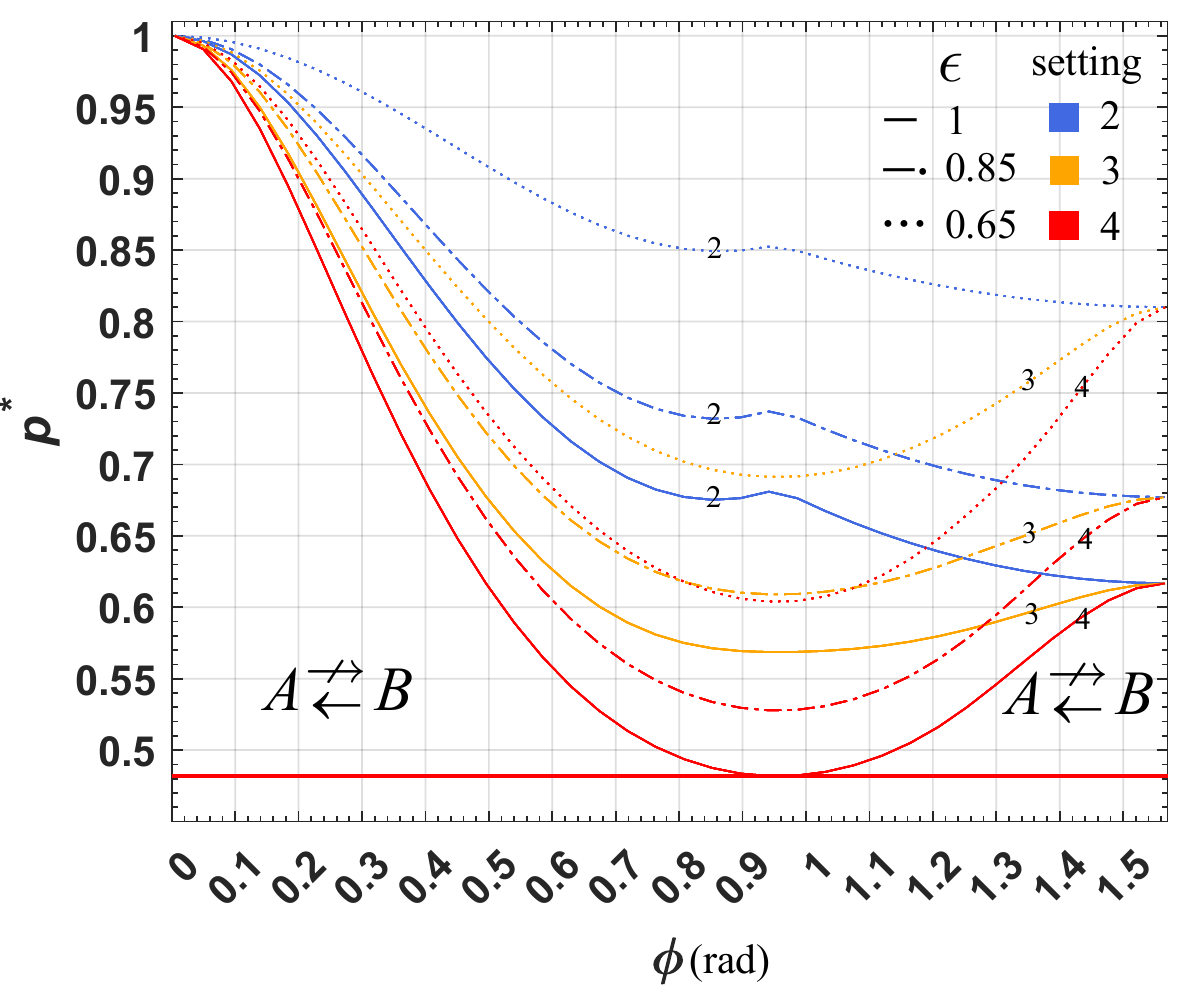}
	\caption{\label{Fig5} Tradeoff relation between the heralding efficiency and the measurement settings for two-qutrit partially entangled state.
	The cross section with $\theta=\pi/4$ is selected as the representative, for there is a two-way steerable point.
	Results of two-, three-, and four-setting measurements are denoted in blue, orange, and red respectively (additional labels on top of each curve are for grayscale version).
	Results of $\epsilon=1,0.85,0.65$ are denoted with solid, dash-dotted and dotted lines respectively.
	The threshold of Bob steering Alice is also presented at the lowest forming a straight line.
	}
\end{figure}
	It is worth noting that the normalization condition is automatically satisfied with $\text{Tr}\sum_\lambda \sigma_\lambda=1$, should the LHS model $\{\sigma^{\text{LHS}}_{a|x}\}$ can fully reproduce the given assemblage $\{\sigma_{a|x}\}$.
	Otherwise the LHS model contributes partially to the reconstruction of $\{\sigma_{a|x}\}$ as shown in Eq.~\eqref{propotion}.
	Thus we can straightforwardly write down the SDP to solve the steering weight with losses counted~\cite{SWLoss}, namely, to maximize $\text{Tr}\sum_\lambda \sigma_\lambda$ over $\{\sigma_{\lambda}\}$, subject to the constraints in Eq.~\eqref{prisiglam}.
	
	Note here we assume that the probability of loss, or equivalently the heralding efficiency $\epsilon$ is independent of the measurement $x$ that Alice performs, which means the action of discarding is completely random and not specified to particular measurement. 
	However, if the efficiency $\epsilon$ is dependent on the choice of measurement, one can simply replace $\epsilon$ with $\epsilon(x)$ in the above expressions.
	
\subsection{Examples}
	Indeed, there are several analytical works have discussed the steering test with losses are taken into account in two- and higher-dimensional systems~\cite{evans2013Losstolerant,skrzypczyk2015Losstolerant}.
	In Fig.~\ref{Fig4}, we present the comparison between our loss-counted model and the known analytical bounds on the cutoff value of visibility of the $d$-dimensional isotropic state versus loss.
	The circles denote our numerical results for the specific dimension, measurement setting and heralding efficiency.
	The authors in Ref.~\cite{skrzypczyk2015Losstolerant} derived bounds for dimension up to five given $d+1$ MUB, and the bounds are denoted in blue, orange, red and magenta respectively.
	
	We can see clearly that our results indicate stronger loss-tolerance compared to the known bounds when loss is low, and are approaching to the theoretical limit along with the increasing of loss.
	This is reasonable since the bounds in Ref.~\cite{skrzypczyk2015Losstolerant} consider the asymptotic behavior of the cutoff value given $d$ and $\epsilon$,
	which leads to the inexact $p^*$ when loss is low. 
	However, we note that a set of refined loss-tolerant steering inequalities from Ref.~\cite{skrzypczyk2015Losstolerant} can probably reproduce the tighter bound.
	
	On the other hand, our results fully reproduced the optimal bounds (denoted in gray) derived in Ref.~\cite{evans2014Optimal} for two-qubit case with two- and three-setting measurements, as indicated in the figure.
	It is worth noting that though the examples are tested with $d+1$ MUB in our work, the model we developed is applicable to any finite measurement settings, such as Platonic solid measurements adopted in Ref.~\cite{evans2014Optimal}. In fact, the two- and three-setting MUB coincide with the square and octahedron measurements respectively.

	It is important to note that the analytical bounds were derived specifically for symmetric states, and our method is applicable to any input state, which is shown as follows.
	In Fig.~\ref{Fig5} we present the loss-counted one-way steering demonstration of two-qutrit partially entangled states, using two-, three- and four-setting MUB.
	For each setting of MUB, we test $\epsilon=1,0.85,0.65$ respectively.
	Note here we only select the cross section with $\theta=\pi/4$ and we only consider the loss on Alice's side, and Bob's device is assumed to be well-characterized---still, Bob cannot steer Alice if the state is unsteerable from his side.
	
	One can see clearly the tradeoff relation between the detection efficiencies and the measurement settings on the steering threshold from Alice to Bob for the partially entangled states.
	Specifically, we see the state is two-way steerable at the point of $\phi=0.9553$ (within numerical precision~\cite{MATLAB}) with four measurement settings, and becomes completely one-way steerable when losses are counted.
	On the other hand, given certain amount of losses, increasing the number of measurement settings can effectively improve the loss-tolerance of Alice to steer Bob, thus giving a larger area of two-way steering.

\section{Conclusion}\label{V}
	In this work, we have introduced a family of high-dimensional one-way steerable states, and taking the three-dimensional case for example, we have fully characterized the parameter regimes of one-way steering, even with respect to general measurements.
	It is worth noting that the validation of one-way steerability of the proposed states for higher-dimensional case is not presented in this work, but a general numerical method is provided---with the dimension is of (relatively small) prime-power number, if adopting the MUB shrinking factor.
	Beyond merely the analysis on the ideal case, we devised an alternative interpretation (which is the variation of the non-click-event interpretation) accounting for the effect of experimental loss and accordingly developed the loss-counted version of steering weight measure.
	We tested our loss-counted model and compared it with known results of symmetric states.
	
	More importantly, our model applies to asymmetric input states.
	Thus we have demonstrated quantitatively the tradeoff relation between the experimental losses and measurement settings for two-qutrit one-way steering.
	The results of this work sheds a light on the experimental demonstrations of high-dimensional one-way steering,
	and on the corresponding quantum cryptography in the one-sided device-independent scenario, where the quantification of loss is crucial.
	 
\section*{Acknowledgments}
	We are grateful to Travis Baker and Howard Wiseman for useful discussions, and especially Chau Nguyen for helpful advice and elaborate reviews on the manuscript.
	We gratefully thank Paul Skrzypczyk for useful discussions on non-steerability certification and helpful advice on the steering weight code.
	This work was supported by the National Natural Science Foundation of China through Grant No.~12105010.
	Q.Z. acknowledges support of The International Postdoctoral Exchange Fellowship Program by China Postdoctoral Science Foundation (Grant No. 20190096).
	

\begin{thebibliography}{49}
\bibitem [{Fur()}]{FurtherExplain}%
\BibitemOpen
\href@noop {} {}\bibinfo {note} {It should be noted that the asymmetric
definition of EPR-steering does not necessarily correspond to an asymmetric
state. It is even possible to study individually the assemblages that the
steering party sends to the steered party regardless of the original state
they may share. In this case the asymmetry of steering is implied by the
directionality that has been already designated.}\BibitemShut {Stop}%
\end{thebibliography}

%

\clearpage
\appendix

\section{One-way steering of two-qutrit partially entangled states with two- and three-setting MUB }\label{S:I}
\begin{figure}[htbp]
	\includegraphics[width=\columnwidth]{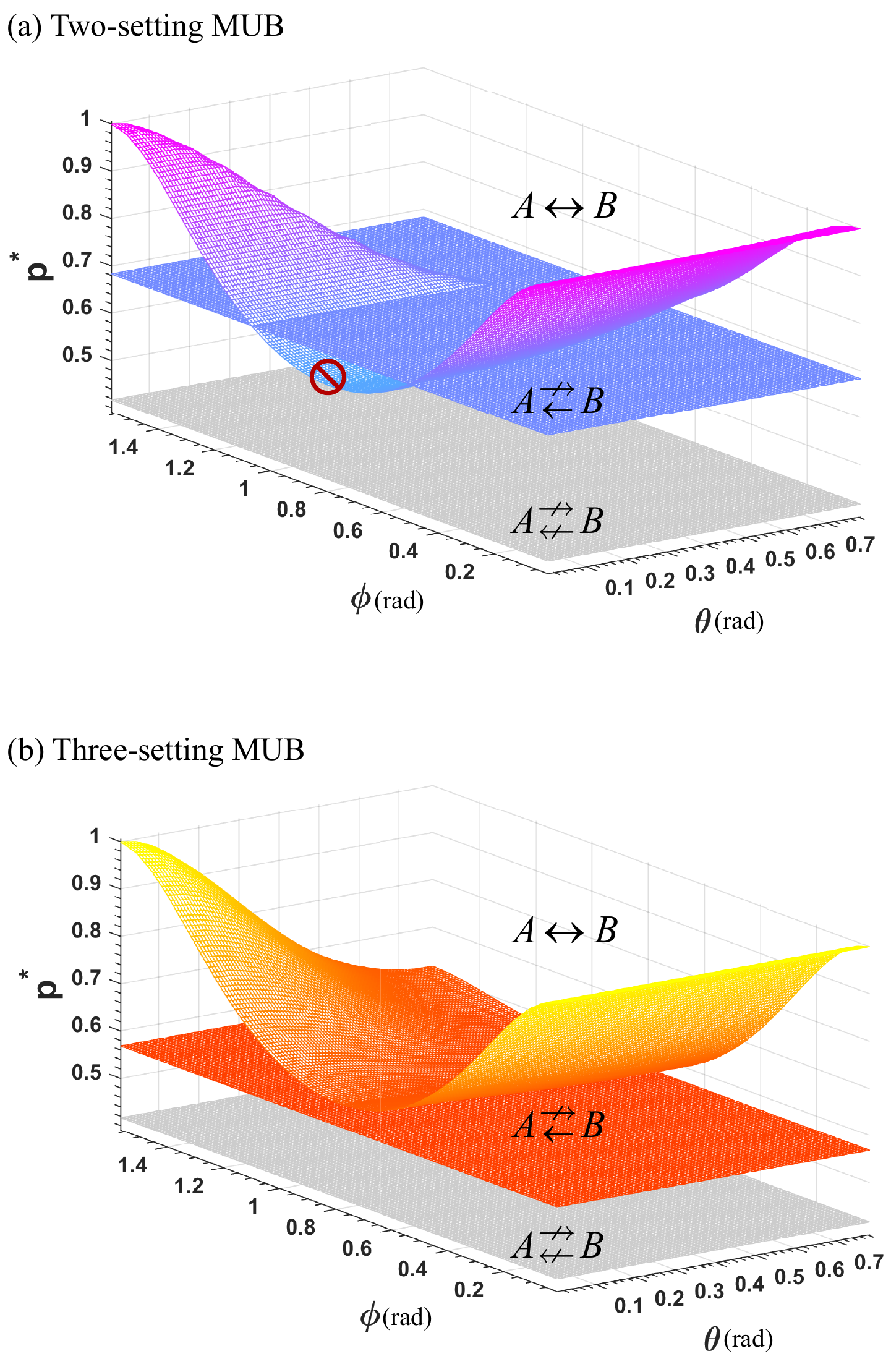}
	\caption{\label{FigS1}  One-way steering demonstration of two-qutrit partially entangled state using two-setting and three-setting mutually unbiased bases.
	}
\end{figure}

	In Fig.~\ref{FigS1} we present the one-way steering parameter space for two- and three-setting MUB.
	It is clear that the overall steering effect with two- and three-settings from both directions are underestimated compared to four-setting case presented in the main text.
	It is worth noting that the one-way steering parameter space in three-setting case is ``shrunk" , but still keeps the similar parameter combination compared to the four-setting case for demonstrating one-way steering. 
	Interestingly, the one-way steering is reversed given two-setting MUB for certain combination of parameter as shown in Fig.~\ref{FigS1}(a), which we see it resulting from the deficiency of two-setting measurements in characterizing three-dimensional system.



\end{document}